\documentclass[useAMS,usenatbib]{mn2e}
\usepackage{graphicx}
\newcommand{\eg}{{e.g. \/}}
\newcommand{\ie}{{i.e. \/}}

\newcommand{\Zsun}{\ensuremath{{\rm Z}_{\odot}}}

\title[Missing Mass in SED-fitting of Unresolved Galaxies]
{Missing Stellar Mass in SED Fitting: Spatially {{Unresolved}} Photometry can Underestimate Galaxy Masses}
\author[R. Sorba and M. Sawicki]{R. Sorba\thanks{E-mail:
rsorba@ap.smu.ca; sawicki@ap.smu.ca} and M. Sawicki\\
Department of Astronomy and Physics and the Institute for Computational Astrophysics, \\Saint Mary's University, 923 Robie Street, Halifax, Nova Scotia, B3H 3C3, Canada}
\begin{document}

\date{Accepted, Received}

\pagerange{\pageref{firstpage}--\pageref{lastpage}} \pubyear{2014}

\maketitle

\label{firstpage}

\begin{abstract}
We fit model spectral energy distributions to each pixel in 67 nearby ($<$$z$$>=0.0057$) galaxies using broadband photometry from the Sloan Digital Sky Survey and GALEX. For each galaxy, we compare the stellar mass derived by summing the mass of each pixel to that found from fitting the entire galaxy treated as an unresolved point source. We find that, while the pixel-by-pixel and unresolved masses of galaxies with low specific star formation rates (such as ellipticals and lenticulars) are in rough agreement, the unresolved mass estimate for star-forming galaxies is systematically lower then the measurement from spatially-resolved photometry. {{The discrepancy is strongly correlated with sSFR, with the highest sSFRs in our sample having masses underestimated by 25\% (0.12 dex) when treated as point sources. We found a simple relation to statistically correct mass estimates derived from unresolved broad-band SED fitting to the resolved mass estimates: $m_{resolved} = m_{unresolved}/(-0.057\log{(sSFR)} + 0.34)$ where sSFR is in units of yr$^{-1}$.}} We study the effect of varying spatial resolution by degrading the image resolution of the largest images and find a sharp decrease in the pixel-by-pixel mass estimate at a physical scale of approximately 3 kpc, which is comparable to spiral arm widths. The effects we observe are consistent with the ``outshining'' idea which posits that the youngest stellar populations mask more massive, older -- and thus fainter -- stellar populations. {{Although the presence of strong dust lanes can also lead to a drastic difference between resolved and unresolved mass estimates (up to 45\% or 0.3 dex) for any individual galaxy, we found that resolving dust does not affect mass estimates on average. The strong correlation between mass discrepancy and sSFR is thus most likely due to the outshining systematic bias.}}
\end{abstract}


\section{Introduction}
\label{sec:intro}

Exactly how a galaxy's stellar content evolves over time is unknown to us. {{In general, the build-up of stars---thought to be a steady stream of star-formation complemented by stochastic bursts triggered by interactions with neighboring galaxies---is halted at certain times by various quenching mechanisms, such as tidal stripping or feedback from \eg active galactic nuclei. While the overall picture is clear, the precise star-formation history (SFH) of any individual galaxy is hidden from us.}} Over the past decade it has become apparent that a galaxy's evolution is {{in large part described}} by its stellar mass. Several tight correlations have been found linking the stellar mass of a galaxy to a wide range of properties.  A galaxy's brightness and size \citep{Trujillo2006, Williams2010}, rotational velocity or velocity dispersion \citep{Faber1976, Tully1977}, star formation rate \citep[SFR;][]{Reddy2006} if on the ``main sequence'' \citep{Daddi2007, Sawicki2012a}, metallicity \citep{Savaglio2005, Mannucci2010}, and mean stellar age \citep{Gallazzi2005}, are all well correlated with the mass of a galaxy's stellar content. 

{{Because a galaxy's stellar mass plays such a vital role in our understanding of galaxy evolution, it}} behooves us to understand the estimating methods available as well as their limitations. Stellar population synthesis (SPS) modeling \citep[for a review]{Bruzual1993, BC03, Maraston2005, Conroy2009, Conroy2013} is a tool used nearly universally to estimate stellar masses in large surveys and at high redshift \citep[\eg][]{Sawicki1998, Kauffmann2003, PerezGonzalez2008, Marchesini2009, Ilbert2010, Sorba2010, Maraston2013}. At its most basic, an SPS model can be created by combining all the light from stellar models in a simple stellar population (SSP)---defined as a group of stars all born at the same time, with the same metallicity, and with a mass distribution described by a chosen initial mass function (IMF)---and tracking how the combined light changes as the stars evolve with time. Composite stellar population (CSP) models can then be created by adding different SSPs together as a function of time determined by some fiducial SFH. The stellar mass is determined by fitting a SPS model grid to observed broad-band photometry and scaling the stellar mass of the best-fitting model. 

The uncertainties in the stellar mass estimate derived from {{broad-band}} SPS fitting can be large, up to a factor of 0.3 in dex when all uncertainties in model assumptions are taken into account \citep{Conroy2009}.  This large variance does not, of course, account for any assumptions that we are not aware of. For example, \citet{Maraston2006} found that differing treatment of the thermally pulsating asymptotic giant branch (TP-AGB) stage of stellar evolution systematically decreased the predicted stellar mass of galaxies by a factor of two. Note, however, that there is still some controversy over how to correctly treat these stars \citep{Kriek2010}. Another model assumption that systematically affects the derived stellar mass estimates is one's choice of IMF, since the IMF directly affects the mass-to-light ($M_*/L$) ratio of the models. A third systematic bias was first noted by \citet{Sawicki1998} and then emphasized by \citet{Papovich2001} who stated that star-forming galaxies had the potential to hide underlying older stellar populations by outshining them. The fitting procedures preferentially favor matching the large amount of flux coming from the younger stellar population, and could potentially under-estimate the $M_*/L$ ratio by missing the relatively low amount of flux emanating from the older stars (where a majority of the stellar mass could rest). {{Unlike the two previously mentioned biases, the outshining bias is inherent to the broad-band SPS fitting procedure, rather than emerging explicitly from one's choice of models. It is thus perhaps particularly insidious because it cannot be easily corrected for from one study to another. Moreover, because the bias would affect star-forming galaxies more strongly than quiescent galaxies, its differential nature could act to either produce spurious relations or mask true ones.}}

{{The outshining bias essentially arises from our lack of knowledge of the SFH of any particular galaxy, and the SFH's influence on the $M_*/L$ ratio. If it were possible to know the precise SFH to use when creating a model SED, then we would know the exact ratio of old to young stars and, hence, would not miss any of their mass when fitting preferentially to the flux emanating from the younger stellar population. However, since the SFH is unknown, its form must be assumed and often parametrized in the model grid to account for many possible differing histories. \citet{Maraston2010} and \citet{Pforr2012} used mock galaxies to test the effects of mismatched SFHs on derived parameters, and found that the stellar masses of star-forming galaxies are typically underestimated, in part due to outshining, by 0.3 dex for $z \sim$ 2 galaxies and up to 0.6 dex for lower redshift galaxies. They found that the mass measurement could be better reproduced by using an inverted-$\tau$ SFH (\ie exponentially increasing), which is presumably closer to the true SFH of their galaxies. However, even if the true SFH is present in the model SED grid, unless it is the only SFH present, an outshining bias can still occur. Because the flux of younger stellar populations dominates so thoroughly at optical wavelengths, any two models with roughly similar young stellar components would have roughly an equal chance of being selected as the best-fit model, even if one model had the true SFH and the other a vastly different one. Outshining can thus be regarded as the loss of constraints on the SFH (and thus $M_*/L$ ratio) because the details are hidden behind bright, young stars. \citet{Gallazzi2009} explicitly examined the uncertainties in $M_*/L$ in the case where the true SFH is present in the model library, with the assumption that the redshift is known and neglecting dust corrections. They were able to quantify biases from mismatched SFHs and in the prior distribution of SFHs compared to the true distribution, which they found could be as high as 0.1 dex.}}

The extent of any bias from {{outshining in our observations of low redshift galaxies}} has been hinted at, but not rigorously determined. \citet{Drory2004} compared masses derived from spectra to photometric masses and found that for objects with strong $H\alpha$ equivalent widths (\ie star-forming), the photometric mass was systematically lower than the spectroscopic mass by up to 0.15 dex. \citet{Zibetti2009} fit SPS models to individual pixels in nine SINGS galaxies, and compared the resolved and unresolved mass estimates. They also found that the unresolved mass estimate (\ie fitting all the light from the galaxy at once as if it were a point source) could be underestimated compared to the resolved mass by up to 40\% (0.15 dex). However, they attribute this effect to unresolved dust lanes, rather than star-forming regions. \citet{Taylor2010}, on the other hand, compared stellar masses with dynamical masses and placed an upper limit on any differential bias in stellar mass estimates at less than 0.12 dex. It is clear that a systematic bias is present, although the large scatter in any individual mass measurement make it difficult to put an exact value on the offset due to outshining. In order to have the best understanding of stellar mass estimates, it is crucial to better constrain this bias. The large statistical variance is stellar mass estimates necessitate the use of a large number of objects to penetrate through the noise.

With this in mind, we follow the approach of \citet{Zibetti2009} and compare spatially resolved versus unresolved mass estimates, as this method only requires broad-band photometry which is readily available from a myriad of different surveys. We set out to constrain the average systematic offset from fitting broad band photometry of star-forming galaxies to SPS models. {{We stress that in this work we are in no way trying to determine the actual amount of stellar material in galaxies, but are instead focused on a comparative study between the pixel-by-pixel and unresolved broad-band SED fitting mass estimates. We are solely interested in the effects of outshining as an inherent procedural bias.}} In Section \ref{sec:method} we describe our data and pixel-by-pixel (PXP) SPS fitting procedure, and in Section \ref{sec:results} we discuss the resulting comparisons between resolved and unresolved stellar mass estimates. Throughout this work we assume a WMAP7 \citep{Komatsu2011} flat $\Lambda$CDM cosmology ($H_0 = 70.4$ km s$^{-1}$ Mpc$^{-1}$, $\Omega_M = 0.272$) and use the AB magnitude system.


\section{Method}
\label{sec:method}

\subsection{Data}
\label{sec:data}

We used publicly available $u, g, r, i, z$ data from the Sloan Digital Sky Survey (SDSS) Data Release 10 \citep{Eisenstein2011} supplemented with publicly available $NUV$ data from the GALEX Ultraviolet Atlas \citep[GUA; ][]{GildePaz2007}, which also provides measurements of the spectroscopic redshift, major and minor diameter, and morphological classification for each galaxy. Although $FUV$ data were also available from GALEX, we found that their signal-to-noise (S/N) ratio was typically too low to get accurate photometry in a single pixel. These six bandpasses should provide more than adequate wavelength coverage to derive stellar masses from SPS fitting, which could be done using only the five SDSS filters\footnote{{We compared stellar mass estimates derived using only the five SDSS filters with those made with the additional NUV bandpass. We found that the inclusion of the NUV had a slight effect on mass estimates, increasing them by approximately 7\% on average, although with large scatter. However, this average difference was present in both the resolved and unresolved mass estimates, and so does not qualitatively affect our conclusions. We postulate that the observed mass difference arises from the NUV limiting the amount of unrealistic extinction available to the fit. That is to say, we found the pixels (and galaxies) were nearly universally fit with higher extinction models when the NUV data was not included. The greater amount of extinction for the same amount of flux led to greater mass estimates when the NUV data was not included.}} \citep[\eg][]{Maraston2013}. The $NUV$ bandpass {{adds a second measurement below the 4000\AA\ break and provides}} additional information about the young stellar population and extinction. It acts to break some of the degeneracies inherent in the models (see Figure \ref{fig:ccplot}).

Starting with the approximately 1000 objects in the GUA which contains all galaxies larger than one arcminute in diameter observed with GALEX, we cross-correlated with the SDSS frames using the bulk image search functionality of the Sloan website. We then visually inspected each matching frame and removed any galaxies that lay too close to the border of the SDSS frame. This step ensures that any edge effects, whether from the raw photometry or due to convolution (see below), will not affect a pixel's final colors. Additionally, we removed any galaxies that spanned the entire SDSS frame since there would be no way to get an accurate background measurement for that frame. The remaining galaxies were processed following the procedure in Section \ref{sec:processing} and color images were made from the resulting aligned and psf-matched images. The color images were again visually inspected and any galaxies with obvious artifacts (\eg excessive bleeding from foreground stars) were discarded. In total these criteria left us with 67 galaxies including eight elliptical, one lenticular, and four irregular galaxies, with the remainder being some form of spiral galaxy.  

\subsection{Processing}
\label{sec:processing}

For each galaxy, we downloaded the appropriate SDSS frame, mask, and point spread function (PSF) which was read at the center of the galaxy using the read\_PSF software package provided on the SDSS website.\footnote{https://www.sdss3.org/dr10/algorithms/read\_psf.php} The GALEX raw count, high resolution relative response, and flag images were downloaded. We use the GALEX PSF available on the Galex website.\footnote{http://www.galex.caltech.edu/researcher/techdoc-ch5.html} We created a square cut-out of the GALEX  images centered on the galaxy with sides equal to 1.5 times the galaxy's major diameter. Although background subtracted intensity images are provided, we found through visual inspection of the sky background images that the provided background maps tended to overestimate the background of extended galaxies, likely due to diffuse radiation from the galaxy \citep[see also][]{Zou2011}. We thus derived our own background estimates in the raw count cut-out by first masking the galaxy and then performing a sigma-clipping function in 50 pixel by 50 pixel square regions and fitting a plane to the background estimates in each region. Since counts from $NUV$ photons are few, {{the estimated background value in each step of the sigma-clipping algorithm was set to be equal to the variance (rather than the mean) of the values in each square region. This was done to account for the lack of symmetry in the Poisson distribution when number counts are low; using the mean would overestimate the background value slightly.}} A background subtracted intensity map with units of counts per second per pixel was created using the formula
\begin{equation}
I = (C - B) / R,
\end{equation}
where $C$ represents the raw count map, $B$ the background map, and $R$ the relative response.
We then created a root-mean-square (rms) noise map for the intensity image by 
\begin{equation}
\sigma_I = \sqrt{\sigma_C^2 + \sigma_B^2} / R,
 \end{equation}
 where $\sigma_C$ is the square root of the count image (\ie Poissonian uncertainty) and $\sigma_B$ is the standard deviation of the background determined using our sigma-clipping method above.

{{The SDSS images are already background subtracted as part of the SDSS pipeline, but this process over-estimates the background of extended objects \citep{Blanton2005, Bernardi2007}. We therefore reinstated the background subtracted by the SDSS pipeline and performed our own background estimation, similar to what is described above, but using the mean in each square region as the best background estimate}} as allowed by the much larger count numbers. Each SDSS frame had to be converted to data numbers using the provided calibration layer of the fits file. The rms images were then made for the SDSS images using the formula 
\begin{equation}
\sigma_I = \sqrt{x\sigma_C^2 + x\sigma_B^2}
\end{equation}
where $I$ is now in units of nanomaggies, $B$ is again the background map, $x$ is the calibration layer value, and $\sigma_C$, the uncertainty in the data number image, is now given by
\begin{equation}
\sigma_C = \sqrt{C/g + \sigma_{dark}^2}
\end{equation}
where $C/g$ is the data number divided by the gain to yield the Poissonian variance of the counts, and $\sigma_{dark}$ is the Poissonian noise from the dark current.  

Having created background subtracted images with corresponding noise maps in each of the desired bandpasses, we next aligned the SDSS images with the GALEX cut-out and resampled the SDSS pixel scale to the GALEX pixel scale (1.5 arcsec per pixel) using the SWARP software package \citep{Bertin2002}. Although losing spatial resolution seems contrary to our purpose, the reasons for degrading the SDSS pixel scale are threefold: 1) Artificially chopping up a NUV pixel into smaller pixels may not result in realistic spatial distribution of the NUV photons (even if interpolation is used), which would lead to erroneous colours in each individual pixel. 2) Combining the SDSS pixels together leads to a higher S/N ratio which is crucial when striving for accurate photometry on a pixel-by-pixel basis. 3) Having fewer pixels speeds up computation time of all future steps immensely. 

We also performed a re-alignment and resampling of the SDSS PSF by placing the PSF at the same location in the frame as the center of the galaxy and using the same parameters for SWARP. It is important to have an estimate of the PSF prior to resampling, otherwise one is likely to have an undersampled PSF and encounter problems when trying to estimate it from the resampled image.

The GALEX PSF was the broadest, so to ensure that each pixel contains light from the same physical location it was necessary to find a convolution kernel that would translate each image from its current PSF to the broadest PSF. We found the transformation kernel by taking the ratio of the two PSFs in Fourier space and then inverse-transforming the result. We masked any saturated pixels and foreground stars \citep[found using the 2MASS catalog;][]{Skrutskie2006} and then convolved each SDSS image with the appropriate transformation kernel using scipy's ndimage convolve routine with pixel values set to have zero value outside the image. The uncertainty was propagated by convolving the variance map with the square of the transformation kernel to yield the variance map of the convolved SDSS image.

With the resulting set of background subtracted, aligned, resampled, PSF-matched images we then made photometry catalogs of each pixel with a S/N of at least 5 in every SDSS filter. To make a fair comparison between pixel-by-pixel SED-fitting result and those using spatially unresolved fluxes, we calculated the unresolved magnitudes not from aperture photometry, but by summing the flux from each pixel included in the pixel-by-pixel catalog and then converting to AB magnitude. Doing this is key as it ensures that the same amount of light is being compared in each case.

\subsection{Models}
\label{sec:models}

\begin{figure}
  \includegraphics[width=84mm]{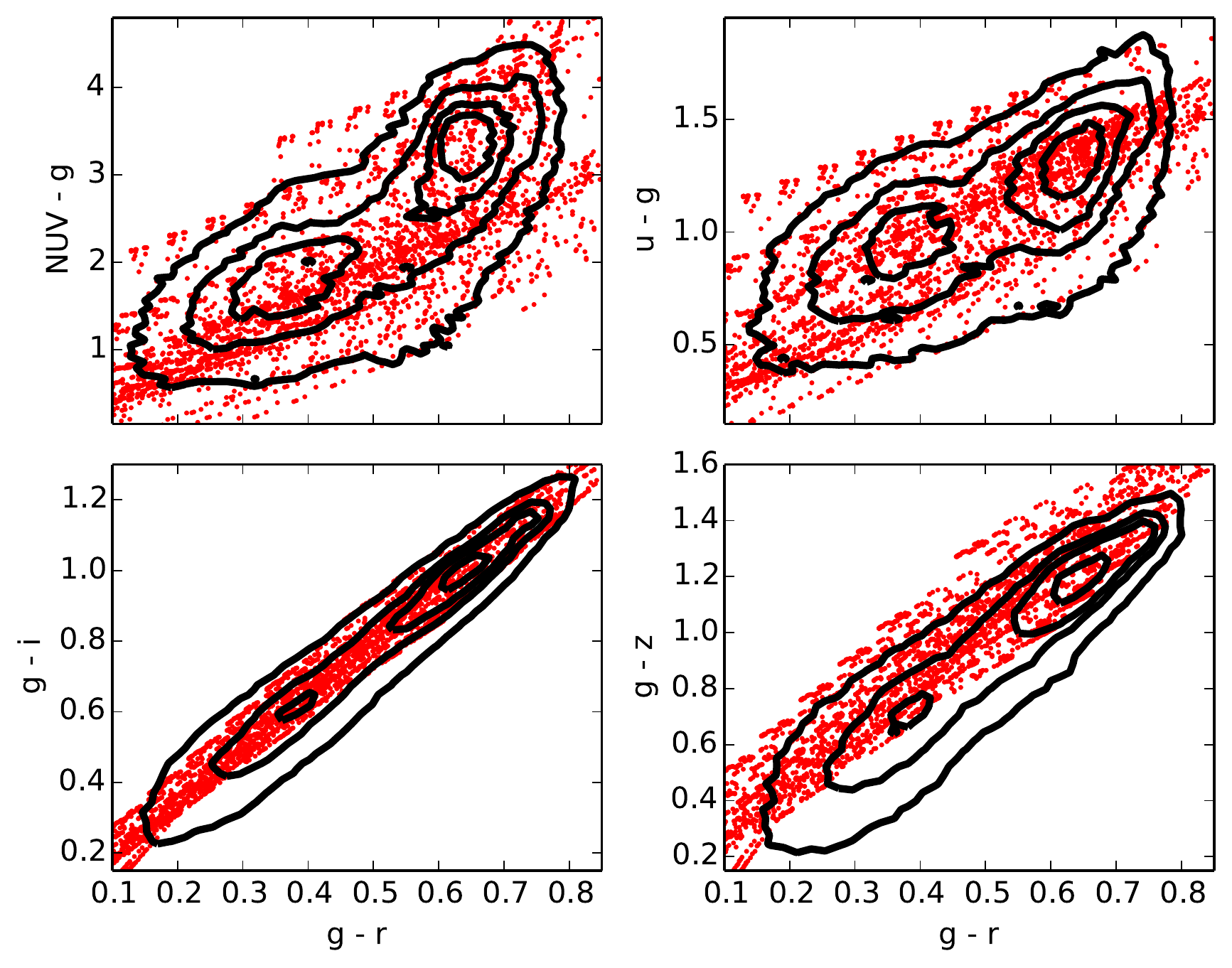}
  \caption{Color-color diagrams showing how our model colors compare with the observed photometry. The black lines represent the 10, 33, 66, and 90\% contour levels of the histogram of the brightest  ($i$-band) quartile of pixels for all galaxies. The total number of pixels binned is 187445. The red points are where our models lie in color-color space for a galaxy at $z = 0.0057$ which is the average redshift of our galaxy sample. The models generally show good agreement with the observations, tending to lie on the locus in each sub-plot.}
  \label{fig:ccplot}
\end{figure}

We fit the broadband photometry from the six SDSS and GALEX bandpasses to model spectral energy distributions (SEDs) created using the FSPS population synthesis code \citep{Conroy2009, Conroy2010}. Our model set is comprised of two component burst models similar to those used in \citet{Noll2009} where there is an underlying older stellar population combined with a more recent burst of star formation. In principle, each of the two components (young and old) can have its own SFH, age (\ie the time since the burst was initiated), IMF, and metallicity history, as well as various parameters controlling how dust absorbs and re-emits light. Here, we describe our chosen model parameters, which we found to adequately represent our photometric observations (listed in Table \ref{tab:modelProps}).

\begin{table}

\caption{\label{tab:modelProps} Parameters used in two-component model SEDs}
\begin{tabular}{@{}lc}
\hline
Symbol & Value(s) \\
\hline
$Z_{old}$ & 0.0016 ($\sim$0.084 \Zsun)\\
$Z_{young}$ & 0.019 (\Zsun)\\
$\tau_{old}$ & [0.1, 0.25, 0.625, 1.25, 2.5, 5, 10]  \\
$\tau_{young}$ & 0.1 \\
$t_{old}$ & 10 \\
$t_{young}$ & [6, 6.5, 7, 7.5, 7.75, 8, 8.25, 8.5, 8.75, 9, 9.25, 9.5] \\
$f_{young}$ & [0, $10^{-4}$, $10^{-3}$, 0.01, 0.1, 0.25, 0.5, 0.75, 0.9, 0.99] \\
$E(B-V)$ & [0, 0.05, 0.1, 0.15, 0.2, 0.25, 0.3, 0.35, 0.4, 0.5] \\
\hline
\end{tabular}
\medskip
The ages are in units of log years.
\end{table}

\begin{figure*}
  \includegraphics[width=\textwidth]{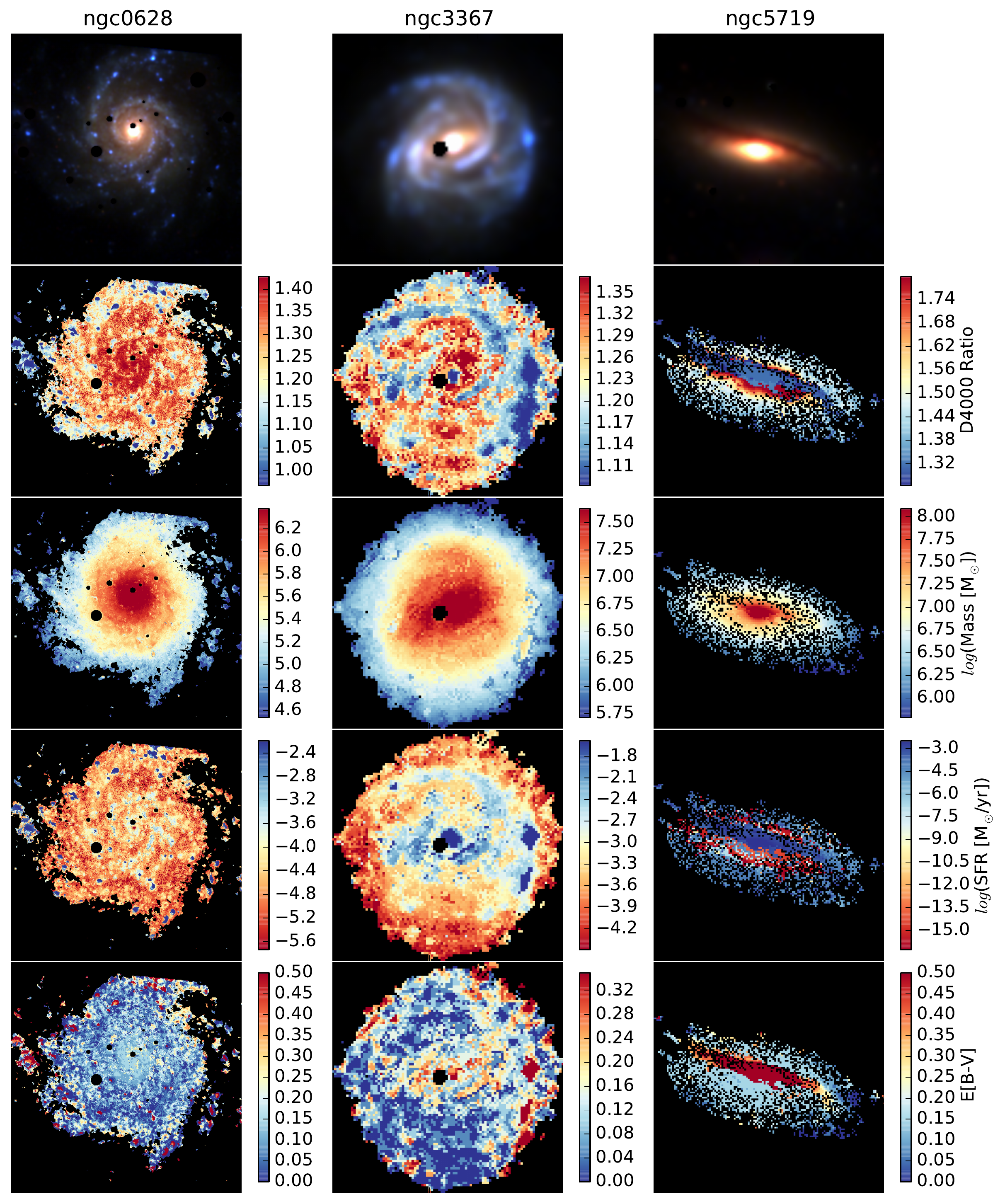}
  \caption{Maps showing (from top to bottom) false color $ugi$, 4000\AA break strength, stellar mass, SFR, and extinction in each pixel for three illustrative galaxies from our sample of 67. Areas of low S/N as well as those containing foreground Galactic stars are masked out in black. Maps for the remaining galaxies can be found in an online supplement.}
  \label{fig:maps}
\end{figure*}

We parametrized the SFH of the older stellar component as a delayed (sometimes referred to as extended) exponential decay (\ie of the form $\psi(t) \sim te^{-t/\tau}$ so that SF at first increases linearly with time) with e-folding time $\tau_{old}$ spaced roughly logarithmically between 0.1 and 10. {{The delayed-$\tau$ models \citep{Lee2010} are a good compromise between the commonly used $\tau$ models and the inverted-$\tau$ models of \citet{Maraston2010}, particularly for low redshift galaxies which have seen a decline in SFR density since $z \sim 2$ \citep{Madau1996, Sawicki1997}. \citet{Simha2014} found that delayed-$\tau$ models were much more successful at reproducing the SFH of simulated galaxies than plain $\tau$ models.}} The SFH of the younger burst is a standard exponential decay with $\tau_{young}$ held constant at 0.1. Each of the young and old components also has an age ($t$). For the older component, this was held constant with star formation starting 10 billion years ago, while the younger burst had onset ages ranging from $10^9$ to $10^6$ years ago. We combined the older and younger components so that the mass fraction of the younger component has a value $f_{young}$ which we allowed to vary between zero and one. {{The two component models used here provide a wide variety of SFHs and are similar in construction to those of \citet{Noll2009} which were found to adequately represent galaxies in the SINGS survey similar to those being studied here.}}

We used a \citet{Chabrier2003} IMF for all models and set the metallicity of the younger component equal to solar metallicity and that of the older component equal to 0.0016 (\ie $\sim$8\% \Zsun). The choice of significantly sub-solar metallicity for the older stellar component was motivated by the simplistic assumption that the overall metallicity of the universe should be lower at earlier times, as well as the results of \citet{Zou2011}, which showed evidence for a double peaked metallicity distribution in NGC 628 with the metal poor component peaking at [Fe/H] $= -1.15$ . Additionally we found that the inclusion of a sub-solar metallicity component yielded a model space that better matched the observed pixel-by-pixel colors (see Figure \ref{fig:ccplot}). Particularly, the $g-z$ color was too red if solar metallicity was used in both stellar components.

For extinction due to dust, we assumed a \citet{Calzetti2000} dust law with $E(B-V)$ values ranging between 0 and 0.5. To save computation time and limit the parameter space we arbitrarily enforced models comprised mainly of older stars to only have $E(B-V)$ values less than 0.3. 

In total our parameter space is defined by four free parameters: $\tau_{old}, t_{young}, f_{young},$ and $E(B-V)$. In Figure \ref{fig:ccplot} we show a comparison between our models at $z = 0.0057$ (the average redshift of our galaxy set) and the pixel-by-pixel photometry for the brightest $i$-band quartile of pixels histogrammed from all galaxies. Although there is a larger spread in the observations than the models due to photometric noise and differing redshifts, our models appear to generally lie on the locus of the observations and adequately span the color-space. It is important to note, however, that since we are doing a comparative study of resolved versus unresolved fitting, it is not imperative that our choice of models actually represent reality to the fullest extent as long as the same set of models is used throughout the study.

In order to determine the best-fitting model to a given galaxy's pixel's broadband photometry we redshifted the model spectrum to the spectroscopic redshift of the galaxy and corrected for dust in the Milky Way \citep{Schlegel1998, Schlafly2011}. The spectrum was then convolved with the broadband filter transmission curves to give model magnitudes in each of the six bandpasses of interest. We used the SEDfit software package \citep{Sawicki2012b} to generate the model magnitudes and then to find the best-fitting model employing a $\chi^2$ minimization routine. In order to determine uncertainties, SEDfit performs a number of iterations for each pixel with the photometry randomly perturbed by selecting a random variable from a Gaussian distribution with standard deviation equal to the uncertainty in the photometry. We performed 300 iterations {{and used the median mass of these iterations as our final mass estimate. This avoids biases in the $\chi^2$ best-fit estimate (see, for example, \citet{Taylor2011}).}}


\section{Results}
\label{sec:results}

In Figure \ref{fig:maps} we show false color images and best-fit model property maps for a subset of our galaxies (the full set {{is available as an}} online supplement). The false color images are made using $u$, $g$, and $i$-band images as the blue, green, and red colors respectively. The property maps show the strength of the D4000 break as a proxy for stellar age \citep[{{since two component models don't have a well defined age;}} see \eg][]{Noll2009}, the stellar mass, the star formation rate (SFR), and the extinction spatial distributions in each galaxy. As expected, regions in the spiral arms that are very bright in $u$-band (\ie flux generated from younger stars) {{are fit to models with weaker D4000 breaks}} and SFRs that are higher than the surrounding population. The mass distribution, however, is relatively smooth, emphasizing that the mass-to-light ratio in each pixel is a function of the stellar population, and that older stars comprise the bulk of a galaxy's total stellar mass while not contributing as significantly to the galaxy's total brightness.

\subsection{Unresolved fitting can underestimate stellar mass}
\label{sec:masscomp}

\begin{figure}
  \includegraphics[width=84mm]{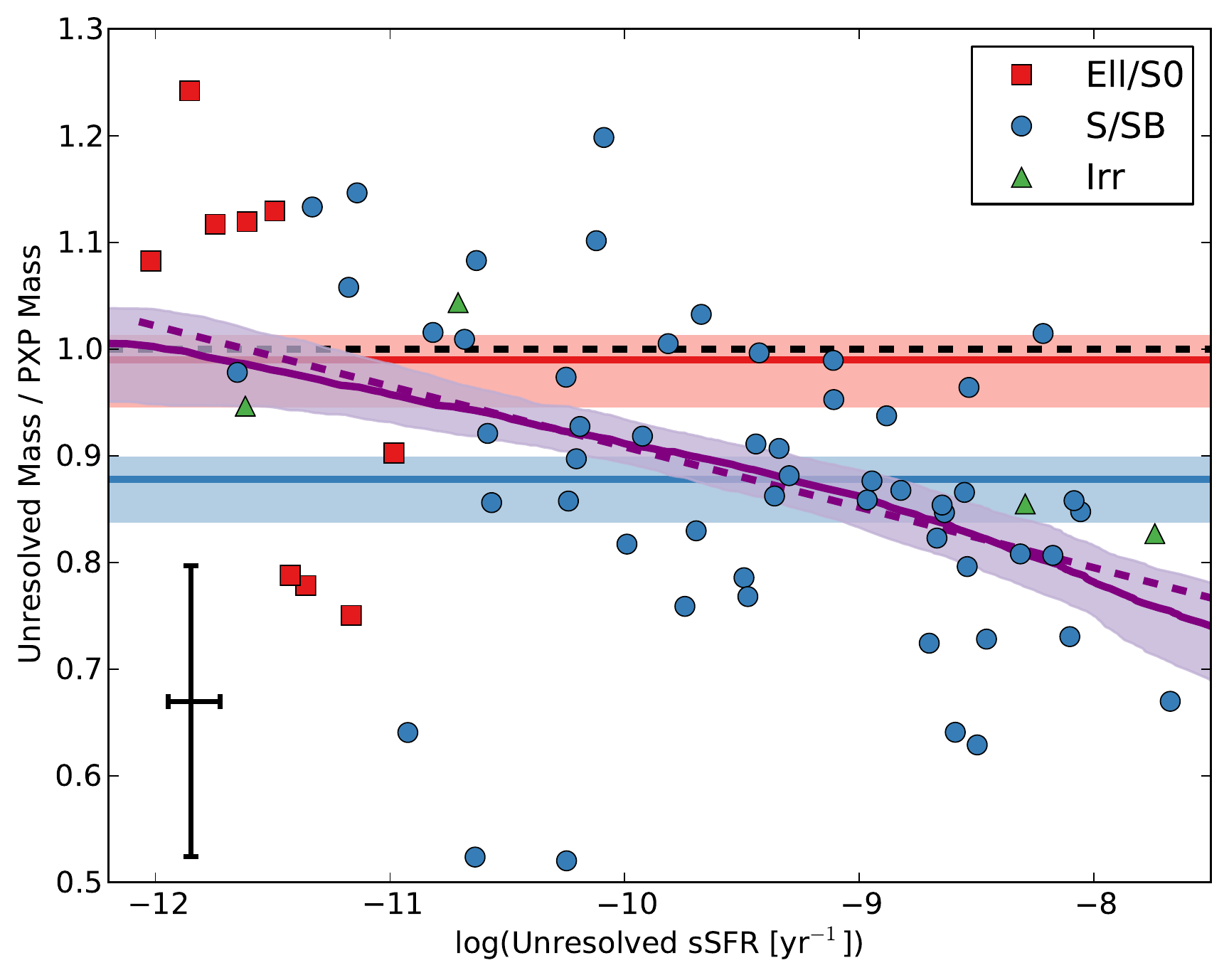}
  \caption{{Ratio of unresolved stellar mass estimate and the PXP stellar mass estimate as a function of sSFR. The dashed horizontal black line represents a one-to-one correspondence. Spiral galaxies (blue points) tend to lie below the one-to-one line, while elliptical galaxies (red points) tend to lie on it (on average). The blue line shows the average ratio of the spiral galaxies indicating $\approx$ 13 percent of the stellar mass is missed by fitting all the light as if the galaxy were an unresolved point source. The red line shows the average ratio for elliptical galaxies, which lies nearly on the one-to-one ratio. The shaded blue and red regions show the 1-$\sigma$ confidence regions for these average ratios as determined from Monte Carlo iterations. The purple solid line and shaded region show the median and 1-$\sigma$ confidence region of fitting a broken linear relationship to all the galaxies in each of the 300 Monte Carlo iterations. The purple dashed line shows a linear fit to the data. A clear trend with sSFR is present. Typical error bars are shown in the lower left corner.}}
  \label{fig:masscomp}
\end{figure}

\begin{figure*}
  \includegraphics[width=\textwidth]{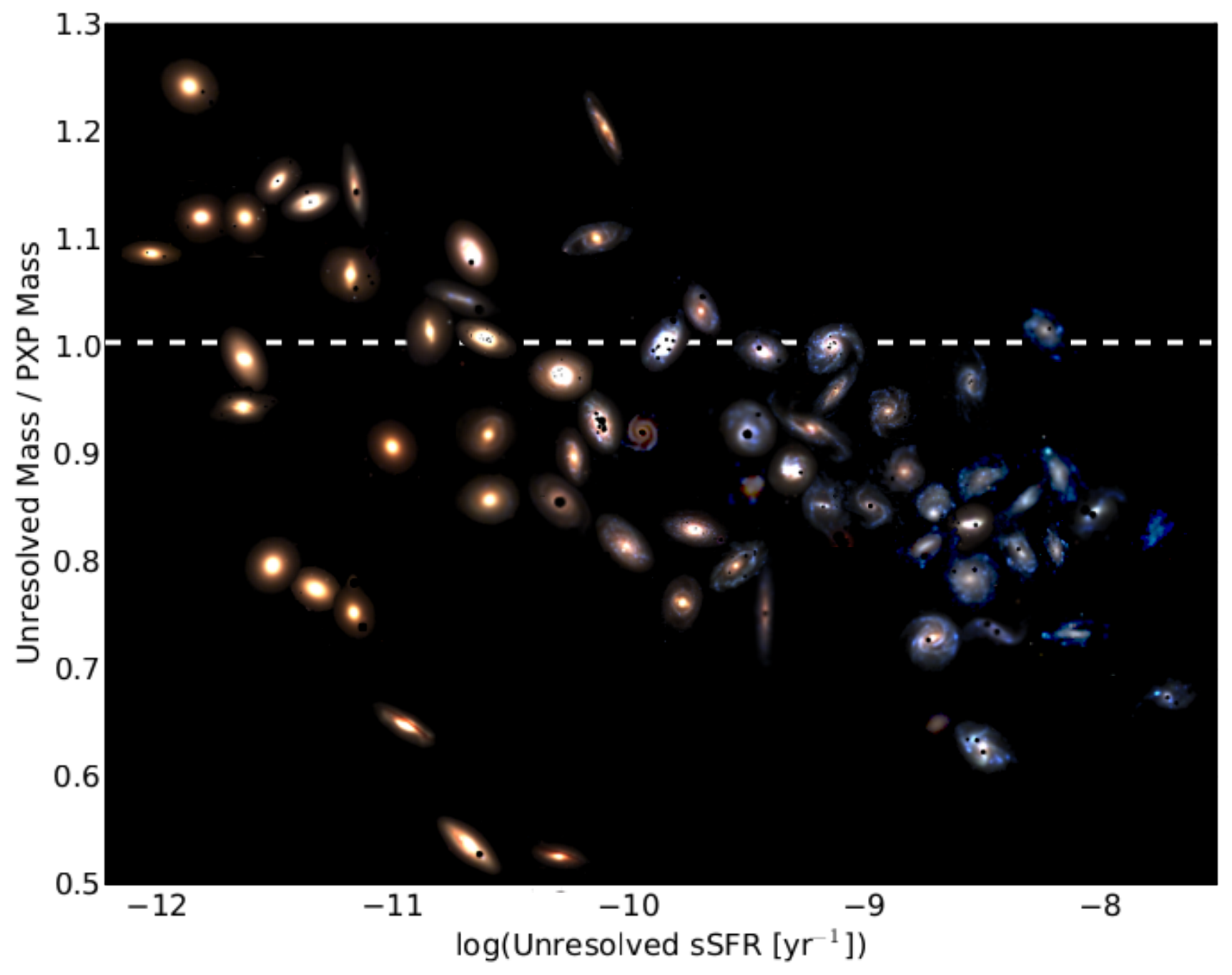}
  \caption{Same as Figure \ref{fig:masscomp}, except showing a false color $ugi$ image for each galaxy. All galaxies have been scaled to be roughly the same size, and some of their positions have been slightly altered for clarity.}
  \label{fig:fc}
\end{figure*}

In order to determine how well unresolved broadband SED fitting can recover the true stellar mass content of a galaxy, we calculated the ratio of the unresolved stellar mass divided by the ``true'' stellar mass (here assumed to be the sum of the stellar mass of the best-fitting model in each pixel). We plot this ratio as a function of unresolved specific star formation rate (sSFR, SFR per unit stellar mass) {{in Figure \ref{fig:masscomp}. We calculated the average unresolved-resolved mass ratio for different morphologies in each of the 300 Monte Carlo iterations and show the median and 1-$\sigma$ spread as horizontal lines in blue for spiral galaxies and red for elliptical and lenticular galaxies.}} Doing so reveals that the stellar mass estimates of elliptical/lenticular galaxies (red squares) tend to have stellar mass estimates that are roughly the same whether resolved or unresolved. On the other hand, spiral galaxies (blue circles) are typically under-estimated by 13\% (0.06 dex) on average when structure is unresolved. {{Moreover, there is a clear decline in the unresolved-resolved mass ratio for the spiral galaxies as sSFR increases. To parametrize this dual nature, we fit the broken linear relation}}
\begin{equation}
f(x) = \left\{
\begin{array}{lr}
b_1 & : x \leq p\\
b_2 + (b_1 - b_2)x/p & : x > p,
\end{array}
\right.
\end{equation} 
{{(which is a horizontal line with intercept $b_1$ up to some point $x=p$ and has some other slope thereafter) to the galaxies in each of the 300 Monte Carlo iterations. The median and 1-$\sigma$ confidence intervals of these fits are shown as the purple solid line and shaded region in Figure \ref{fig:masscomp}. In all Monte Carlo instances, there is a clear trend of increasing average mass discrepancy with increasing sSFR, with unresolved stellar mass estimates underestimated by 25\% (0.12 dex) at the highest specific star formation rates in our set of galaxies. Put another way, to correct the unresolved stellar mass estimate to the pxp stellar mass estimate for these high sSFR, one would have to add 33\% to the unresolved mass, which is much more significant than adding the statistical average correction of 15\% (0.06 dex).}} 

{{Since the break point of the piecewise functions all exist at the lowest extremes of our measured sSFRs, separating the data into two regimes may be an unnecessary complication. For simplicity, we fit a simple linear relation to our median Monte Carlo mass ratios (shown as the purple dashed line in Figure \ref{fig:masscomp}). This line lies within the shaded 1-$\sigma$ confidence region, and provides a close approximation to the more complex median of the piecewise fit distribution. It also allows for a simple formula to convert an unresolved stellar mass estimate to pixel-by-pixel one based on the unresolved sSFR. The conversion relation is
\begin{equation}
\label{eqn:conversion}
m_{resolved} = \frac{m_{unresolved}}{-0.057\log{(sSFR)} + 0.34},
\end{equation}
where sSFR is in units of $yr^{-1}$. }}

\subsection{Origin of the discrepancy}

\begin{figure}
  \includegraphics[width=84mm]{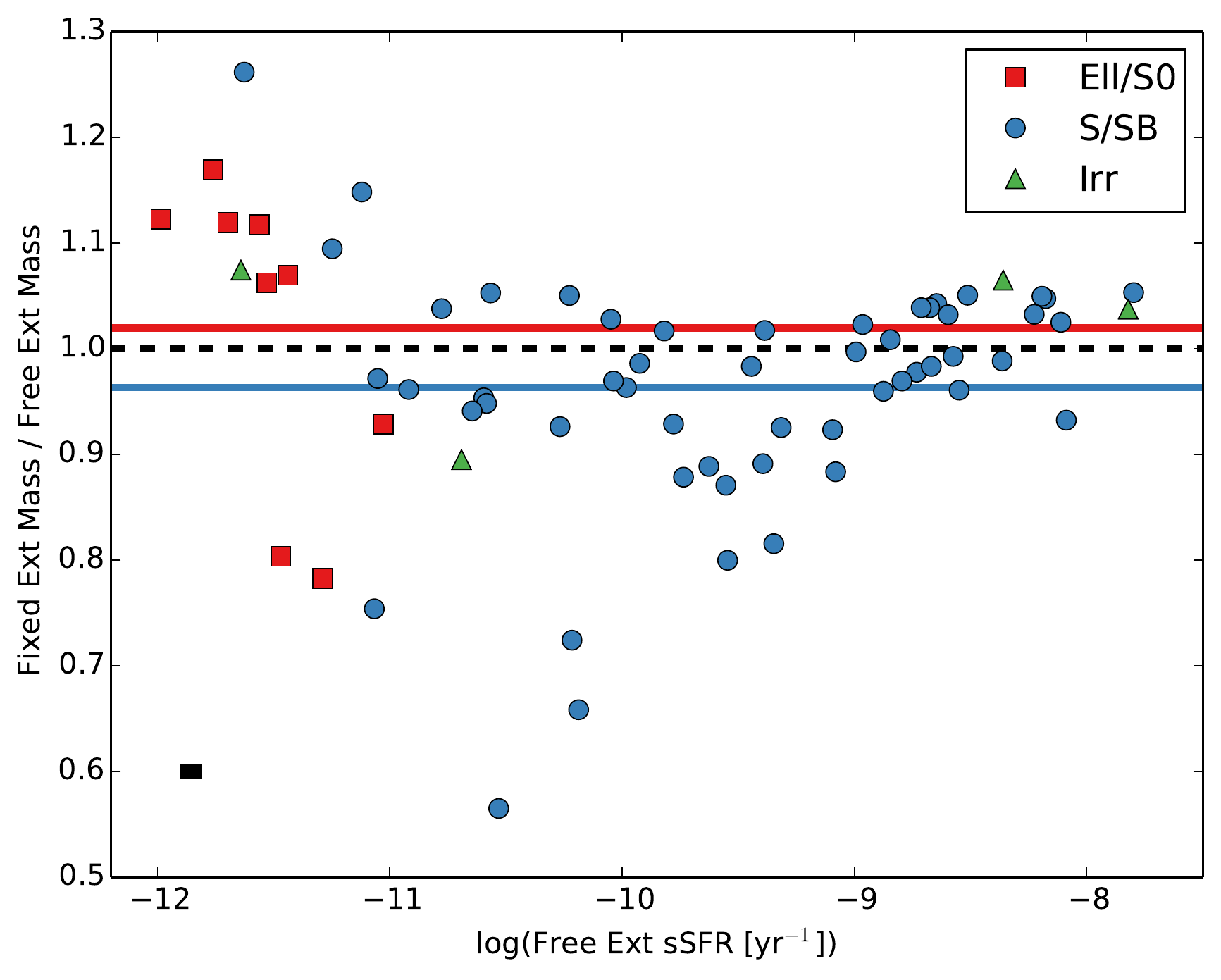}
  \caption{{As in Figure \ref{fig:masscomp}, a mass ratio is plotted versus the sSFR. Blue circles are morphologically classified as spirals, red squares ellipticals, and green triangle irregular galaxies. Here, the vertical axis shows the ratio of the PXP mass found when holding the extinction value fixed to the amount determined by the best-fitting unresolved estimate, to the PXP mass found when extinction is a free parameter. }}
  \label{fig:fixedExt}
\end{figure}

{{In Figure \ref{fig:fc} we superimpose false-color images of each of the galaxies onto their location in Figure \ref{fig:masscomp}. It is apparent that the most egregious outliers from the one-to-one relation (NGC 3190, NGC 4419 and NGC 5719 located in the lower left corner with unresolved-resolved mass ratios less than 0.65, but also edge-on M98 near the top center) all have strong dust lanes with heavy amounts of extinction running through them. To try to determine whether resolving dust or SFR is the dominant driver of the unresolved-resolved mass discrepancy, we refit the pixel-by-pixel photometry but constrained the extinction to be the same as determined from the best fit to the unresolved photometry. The results are shown in Figure \ref{fig:fixedExt} where we plot the ratio between the fixed extinction PXP mass and the free extinction PXP mass versus sSFR. In this figure, the closer the points lie to the same mass ratio as Figure \ref{fig:masscomp}, the more important role dust plays in the unresolved-resolved mass discrepancy, whereas points near the one-to-one line are little affected by resolving dust. In the cases of NGC3190, NGC4419 and NGC5719, the effects of resolving the strong dust lanes accounts for the majority of the difference between unresolved and resolved stellar mass estimates, as they again lie well below the one-to-one relation. We also find that the large majority of the spread in elliptical galaxies and those with low sSFR is due to allowing extinction to be a free parameter from pixel to pixel. Removing this freedom essentially eliminates any difference between resolved and unresolved stellar mass estimates for the low-sSFR galaxies.}}

{{ What is not present in Figure \ref{fig:fixedExt} is any trend with sSFR. Even though the mass difference due to resolving dust can be drastic for any one galaxy, the deviation appears centered on the one-to-one line so that, on average, not resolving dust does not systematically bias mass estimates of large numbers of galaxies. The strong linear trend seen in Figure \ref{fig:masscomp} must then be due to distinguishing the strongly and weakly star-forming regions and, thus, accounting for the outshining bias in broad-band SED fitting.}}

\subsection{The effect of varying spatial resolution}
\label{subsec:resstudy}

\begin{figure}
  \includegraphics[width=84mm]{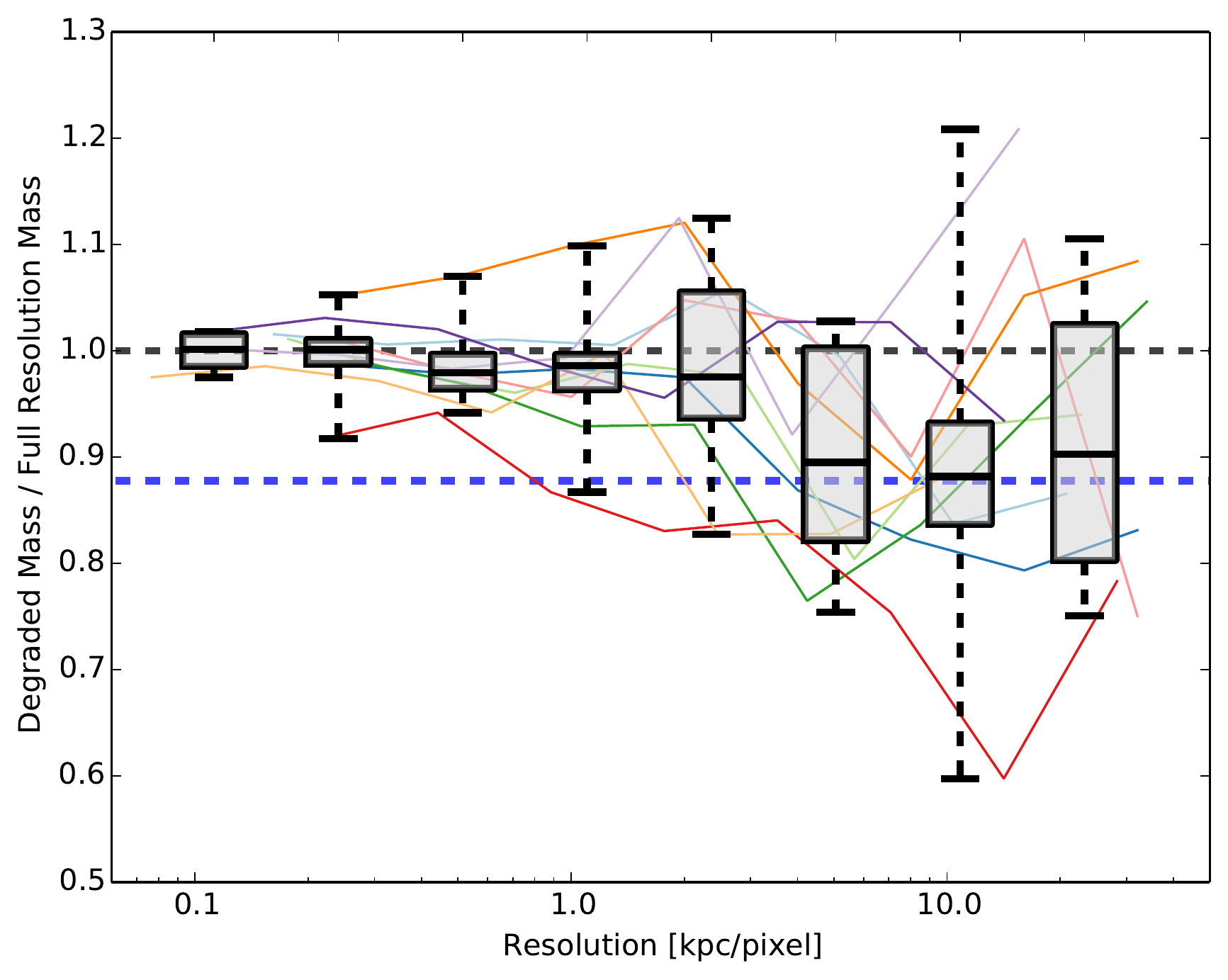}
  \caption{Ratio of degraded resolution PXP stellar mass estimate and the highest resolution PXP stellar mass at various pixel scales for 10 spiral galaxies with the largest area. The boxes and whiskers represent quartiles of the data {{grouped in bins of width $\sim 0.33$ dex in resolution.}} As in Figure \ref{fig:masscomp}, the dashed black line shows the one-to-one correspondence and the dashed blue line the average mass ratio for spiral galaxies. There is a sharp transition in the mass ratio {{near 3 kpc/pixel. The thin solid coloured lines show the individual tracks for each of the 10 galaxies.}}}
  \label{fig:resstudy}
\end{figure}

In an effort to understand how resolution affects the PXP parameter estimations, we artificially degraded the images of the ten largest face-on spiral galaxies by progressively doubling the pixel scale. This process was done for those ten (of our 67) galaxies with large enough areas that after eight degradation steps they were completely unresolved and contained within one or two pixels. Figure \ref{fig:resstudy} shows quartile ranges for the ratio of the degraded resolution PXP stellar mass estimate over the full resolution PXP stellar mass estimate as a function of {{physical}} scale. At the highest resolutions, the mass ratio {{slowly decreases to about 0.97}}, but drops sharply when the pixel scale becomes greater than {{2-3 kpc/pixel}}. The mass ratio then stabilizes near the 13\% under-estimate mark, although with much higher variance in the quartile ranges. It should be made clear that any one galaxy does not display a mass ratio that monotonically decreases with increasing resolution; the variance of mass estimates from SED is too large to show this. It is only by looking at an ensemble of galaxies that we can see on average how the mass estimate behaves as the spatial resolution gets poorer and poorer. 

{{The physical scale of of 3 kpc/resolution element (here being pixel because the artificial degradation makes each pixel larger than the PSF) corresponding with the comparatively steep drop in mass ratio}} is much larger than the size of typical star forming regions in these galaxies \citep[a few hundred parsecs;][]{Efremov1995, Gusev2013}. {{The scale may be more}} commensurate with the width of spiral arms \citep{Kennicutt1982, Seigar1998}. {{To adequately separate young and old stellar populations and thus minimize the effect of outshining bias, it seems necessary to at least be able to resolve any spiral structure in the galaxy. This result may just be an artifact of our  resolution study sample however, as these galaxies all have star forming populations concentrated in spiral arms. Note also that several irregular galaxies shown in Figure \ref{fig:fc} follow the linear trend due to outshining despite the lack of any clear spiral structure. Nevertheless, as spiral structure has been observed to have formed much earlier in the history of the universe than previously thought possible \citep{Law2012}, it is still relevant to know that resolving spiral arms goes a long way toward mitigating any bias due to outshining. Perhaps more generally, one can say that pixel-by-pixel SED fitting can greatly reduce the effects of outshining only when any structure that distinguishes older and younger stellar populations is able to be resolved. In the case of our low redshift sample, this structure is spiral arms with a physical scale of approximately 3 kpc. If this level of resolution is impossible, one can use the conversion relation given in Section \ref{sec:masscomp}.}}

\section{Discussion}

\subsection{{Testing Model Space Dependence}}

{{As discussed in \citet{Gallazzi2009}, the model template grid used constitutes a prior assumption which can bias the mass-to-light ratios determined from broad band colors. In order to test the extent to which our results depend on our model space we performed two tests. Firstly, \citet{Gallazzi2009} found that including a large fraction of bursty models could underestimate the mass-to-light ratio of older, smooth-SFH galaxies. Although we found the unresolved and PXP masses of ellipticals do not differ on average, we nevertheless tested the robustness of our results using a smooth SFH. While it is impossible to exhaustively test every SFH, we repeated our procedure using $\tau$ SFH models and arrived at qualitatively similar results (namely the unresolved mass estimates of galaxies with low sSFRs were no different than the resolved mass estimates on average, whereas star-forming galaxies were underestimated by approximately 0.07 dex on average, although without as strong a dependence on sSFR). From this we conclude that, while our results are not entirely independent of our choice of SFH (one could always construct unrealistic SFHs that would produce vastly different results, say if all star forming models had no underlying older stellar population at all), for any reasonable or commonly adopted SFH our results should be qualitatively similar, with unresolved stellar mass estimates of star-forming galaxies underestimated by somewhere between 0.05 and 0.08 dex on average.}}

\begin{figure}
  \includegraphics[width=84mm]{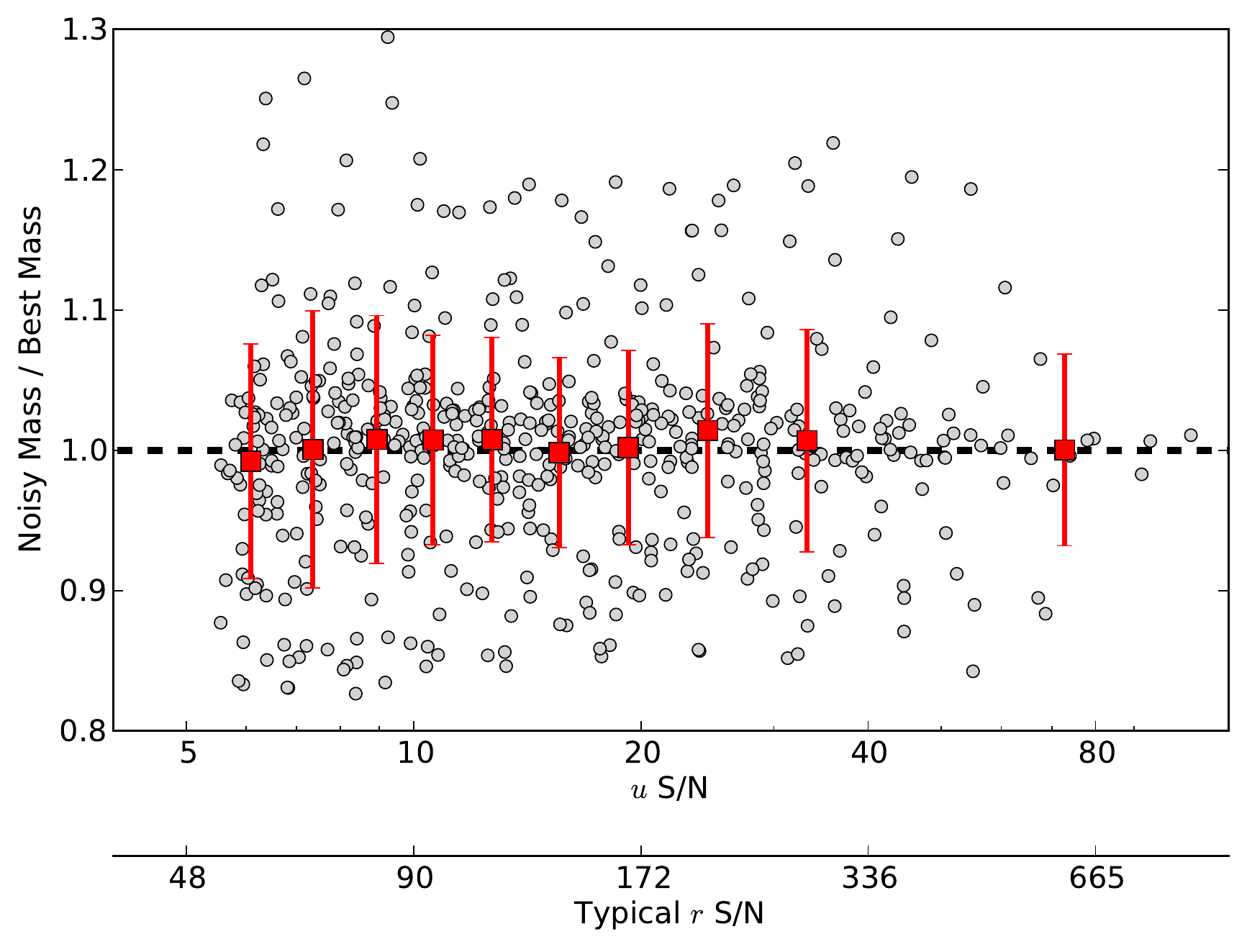}
  \caption{{Ratio of mass estimates from unresolved photometry with artificially increased uncertainties and those with lowest uncertainties as a function of $u$-band $S/N$. The artificially increased uncertainty in each bandpass is determined by taking the $i^{th}$ percentile of the $S/N$ histogram where $i$ is a multiple of 10. The typical $r$-band $S/N$ determined from these percentiles is also shown for reference. The grey points show the mass ratio for each galaxy at each noise percentile. The red squares show the average of the grey points in bins of $1/10^{th}$ of the data. The black dashed line shows the one-to-one correspondence. The mass estimates show no bias with increasing $S/N$ within our cut-off limit of 5.}}
  \label{fig:sntest}
\end{figure}

{{Secondly, as examined in \citet{Gallazzi2009}, the signal-to-noise ratio of the observations combined with the model template prior distribution can lead to biased results. In brief, they found that when the model template set is a mixture of bursty and smooth SFHs with differing $M_*/L$ ratios, good data will constrain the fits to only those models with the proper $M_*/L$ ratio. But, as the amount of noise in the data increases, the fit will be more free to explore the model space, and hence the determined $M_*/L$ ratio will depend on the ratio of bursty to smooth SFHs in the model template set. Since our stated criteria for a pixel to be included in our mass estimation was $S/N$ greater than five in all SDSS bandpasses, this effect could play a prominent role in our results. However, in practice the $S/N$ in $g,r,i$ is much better than that in either $u$ or $z$, and our criteria is essentially equivalent to $S_u/N_u > 5$ (although, in a very small percentage of pixels, the $z$ band is the limiting one). The large $S/N$ in $g,r,i$ give the fits much less freedom than initially implied by our selection criteria. To test this, we took the unresolved photometry (which has extremely good signal to noise) and artificially increased the uncertainty in each bandpass, refitting the median Monte Carlo mass at each step. To realistically approximate the uncertainty growth in each bandpass for our data, for each galaxy we created histograms of the $S/N$ in each bandpass and found the $i^{th}$ percentile where $i$ is a multiple of ten. We converted this to an uncertainty in magnitude for the Monte Carlo fits. Figure \ref{fig:sntest} shows the ratio of the increasingly noisy mass estimate and the unresolved mass estimate as a function of $u$-band $S/N$. The typical $r$-band $S/N$ at this stage is also shown for reference, and is typically 8-10 times that of the $u$-band. The binned average (red squares) shows little deviation from a one-to-one correspondence. There may be a slight downward trend at $S/N_u < 10$, but it is at the 2\% level. Since the low $S/N$ pixels are also the least massive, any bias of this level in their mass determination has a net effect of much less than 1\% on the total PXP mass of the galaxy. Additionally, since the noisy mass would underestimate the true mass, this bias would only work to increase the unresolved-resolved mass discrepancy found above, if only very slightly.}}  

\subsection{Impact on previously derived relations}

\begin{figure}
  \includegraphics[width=84mm]{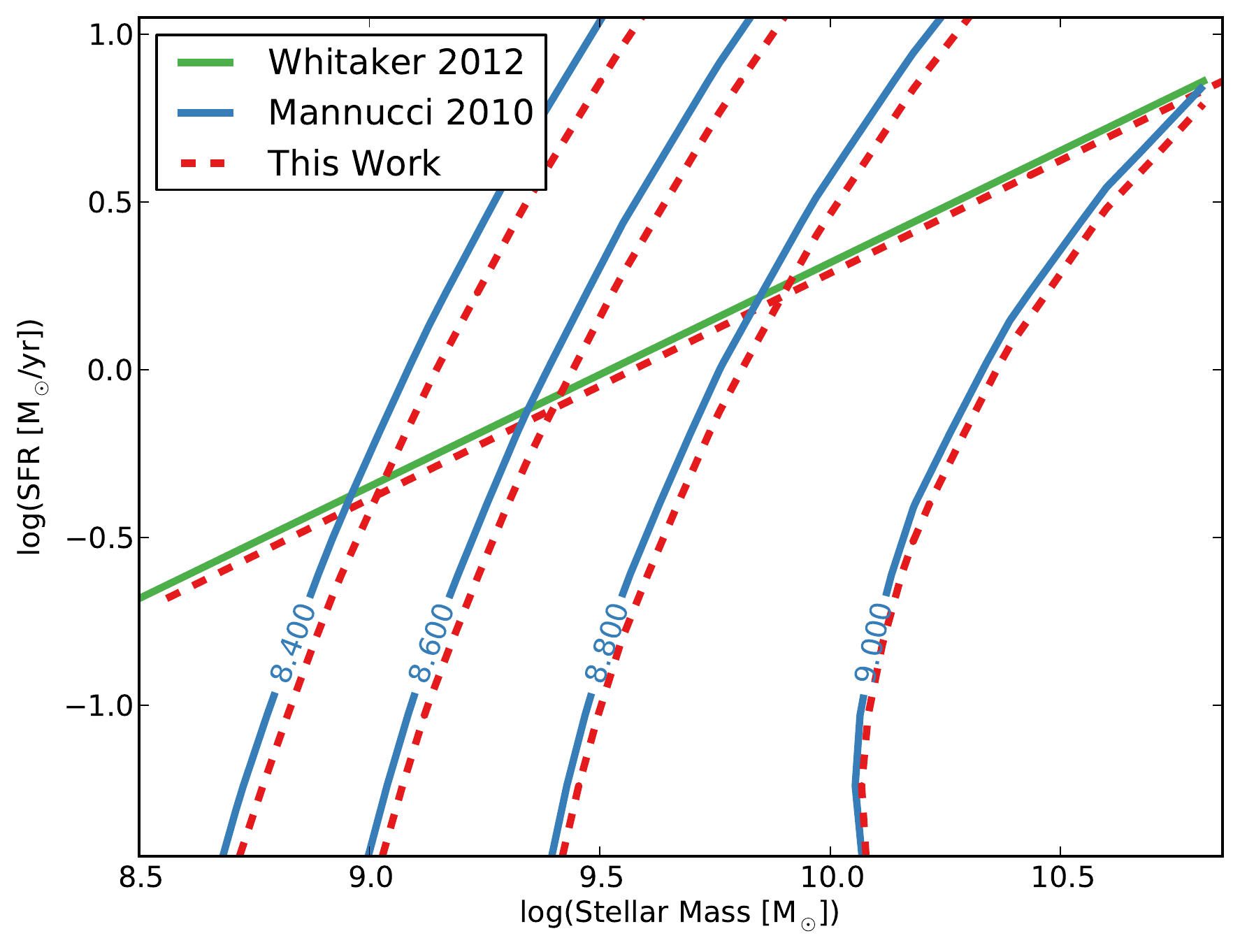}
  \caption{{SFR-$M_*$ relation from \citet{Whitaker2012} in green and isometallicity contours of the $M_*-Z-SFR$ relation from \citet{Mannucci2010} in blue. The numbers on the contours are in units of $12+log(O/H)$. The red dashed lines show corrections to these relations based on this work.}}
  \label{fig:mmsfr}
\end{figure}

{{The average offset of 13\% for star-forming galaxies is equivalent to approximately 0.06 dex. This is not an outstandingly large discrepancy, and any current relationship with stellar mass would not change dramatically if corrected for this offset. However, the differential nature of the offset (affecting higher sSFR galaxies more than those with lower sSFR) can compound the effect given the right circumstances. For instance, plotted in Figure \ref{fig:mmsfr} in green is the star forming main sequence (SFR-$M_*$ relation) from \citet{Whitaker2012} and isometallicities of the $M_*-Z-SFR$ surface from \citet{Mannucci2010} in blue. Dashed red lines show corrections to these relations based on interpolating the purple line in Figure \ref{fig:masscomp} of this work. The shallow slope of the \citet{Whitaker2012} relation in the SFR-$M_*$ log-log plane leads to little difference when corrected for outshining. At best one can say the slope is a little steeper, requiring slightly faster evolution from the flatter relation at higher redshift (assuming no effect of outshing at higher redshift). However, the effects of outshining become more apparent when metallicity dependence is included in the relationship. The $M_*-Z-SFR$ relationship exists throughout the SFR-$M_*$ log-log plane, allowing one to see the larger offsets in high sSFR regions (upper left of Figure \ref{fig:mmsfr}) compared to lower sSFR regions. The steeper slope of the isometallicities also lead to outshining corrections being larger in the full $M_*-Z-SFR$ (though still not extremely large). }}

{{The sSFR dependence of the unresolved-resolved mass difference can manifest itself in other, possibly less obvious ways. For instance, \citet{Patton2013} compare star formation rates in interacting galaxy pairs with a control sample selected to have similar masses as the paired galaxies. Since the SFR is enhanced in the paired galaxies, however, so too is the sSFR. Thus, a larger correction must be applied to the paired galaxy's unresolved mass estimate than must be applied to the control sample, possibly making the control sample not representative. It turns out not to be an issue in this particular case, because the maximum SFR enhancement is approximately a factor of 2, meaning the difference between the enhanced corrected mass and control corrected mass is only 5\%, well within their stated control bounds of 0.1 dex in stellar mass. However, awareness that mass selected samples may not be as homogeneous in mass due to differing sSFRs is still important to acknowledge.}}

{{Even within a single galaxy the sSFR-dependant mass correction may be important to note. Several works \citep[\eg][]{Allen2006, Simard2011, Mendel2014} decompose a galaxy into an inner and outer components (\ie bulge and disc) with SF occuring primarily in the outer regions. The sSFR of the outer region is thus enhanced compared to the inner region, and so a differential correction (perhaps up to 0.1 dex) must be applied when comparing the masses of the inner and outer components. In general, the outshining mass correction should be considered for broad-band SED fitting in any cases with high sSFRs, or when comparing two sets of data with differing sSFRs.}}

\subsection{Outshining at higher redshifts}

How the systematic bias described above might change with redshift is an open question. The sSFR of {{star-forming}} galaxies increases with redshift \citep{Feulner2005}, and so we might expect the outshining effect to become more drastic at earlier times in the universe. {{For example, taken at face value, the relation given by the purple line in Figure \ref{fig:masscomp} implies that galaxies with sSFRs between -8 and -7 $yr^{-1}$ (rates much more common during the era of peak SFR density) would have to be corrected by between 0.1 and 0.2 dex respectively. However, as outshining is a product of incorrect SFHs, galaxies at higher redshift should be less affected because there has been less time for differences in the assumed versus true SFH to accumulate. Put another way, there would be fewer older stars (and hence less mass) hiding behind the bright younger stars simply because there has not been enough time to form a significant older stellar population. See, for example, \citet{Pforr2012} on the effects of assumed SFH as a function of redshift. It is clear that, at some value of redshift, outshining would cease to be an issue as there would not be a significant older population of stars to obscure. In the limiting case, the first burst of stars formed in a galaxy would essentially be a SSP, and thus have no outshining.}}

{{Pixel-by-pixel SED fitting has been performed for high-redshift galaxies at $z \sim 1$ and $z \sim 2$ by \citet{Wuyts2012} using data taken with Wide Field Camera 3 (WFC3) on board the Hubble Space Telescope (HST) as part of the Cosmic Assembly Near-infrared Deep Extragalactic Legacy Survey \citep[CANDELS;][]{Grogin2011, Koekemoer2011}. They found that the unresolved mass estimate from these galaxies was robust, and that there was no discrepancy between the resolved and unresolved mass estimates. Their PSF matched images had a resolution of 0.18 arcseconds FWHM, which corresponds to a physical scale of 1.46 and 1.54 kpc at redshifts 1 and 2 respectively. From our results in Section \ref{subsec:resstudy} (Figure \ref{fig:resstudy}), this resolution should be adequate to detect any discrepancy if it existed at high redshift. However, \citet{Wuyts2012} also further bin pixels to increase signal-to-noise, making their effective resolving power much worse the the PSF FWHM. Indeed, they state that they do find a mass discrepancy of 0.2 dex for $z \sim 2$ star-forming galaxies when no pixel binning is performed, although they attribute this to low signal-to-noise pixels having spurious mass-to-light ratios. How outshining effects mass estimates at high redshift thus remains an open question, one that may not be able to be answered until the advent of the next generation of telescopes such as the James Webb Space Telescope (JWST) or Thirty Meter Telescope (TMT), which will provide roughly 2.5 and 10 times greater resolution than Hubble respectively.}}

\section{Conclusions}
\label{sec:conclusions}

We fit SPS models to {{broad-band fluxes in}} each pixel for 67 nearby galaxies and compared the stellar mass estimates found from summing each pixel's {{median Monte Carlo}} mass with the {{median Monte Carlo}} mass found when the models were fit to all the light from the galaxy at once, as if it were an unresolved point source. 

{{The unresolved mass estimate systematically underestimated the resolved mass estimate, displaying a clear trend of increasing discrepancy with increasing sSFR. Galaxies with low sSFR ($log(sSFR [yr^{-1}]) \sim -12$) were consistent with no mass discrepancy between resolved and unresolved mass estimates, whereas unresolved mass estimates of the highest sSFR galaxies ($log(sSFR [yr^{-1}]) \sim -8$) typically underestimate the mass by 25\% (0.12 dex). We provide a conversion formula (Equation \ref{eqn:conversion}) to correct unresolved mass estimates at low redshift based on a galaxy's sSFR.}} 

{{We found the mass discrepancy's correlation with sSFR to be due to outshining, \ie young stellar populations obscuring older stellar populations behind their bright flux. However, several egregious outliers from this relation were caused by the presence of strong dust lanes. The presence of dust can cause differences between the resolved and unresolved mass estimates of up to 45\% (0.35 dex). Although very significant for individual galaxies, resolving the 2D distribution of dust did not have any affect on average stellar mass estimates.}}      

By artificially degrading the resolution of the largest spiral galaxies, we found that a resolution of at least {{3 kpc}} is needed to significantly reduce the systematic under-estimation in stellar mass due to outshining. {{This scale is commensurate with the width of spiral arms in these galaxies.}} 

Although it is unclear {{how the relation between resolved and unresolved}} stellar mass estimates {{changes}} at higher redshifts, we recommend that {{caution be taken when considering mass estimates of galaxies with high sSFRs that have been derived from broad band SED fitting. Be aware that these masses are likely systematically underestimated by 13-25\% (0.06-0.12 dex) on average due to outshining. The differential nature of this effect should be taken into account when comparing masses of populations with different sSFRs (\eg disc versus bulge, ``mass-matched'' galaxies with a range of SFRs). In the case of any individual galaxy, note that the presence of dust lanes can greatly impact the mass estimate (by a factor of $\sim$2 or 0.3 dex), but that the effects from dust should average out when considering an ensemble.}}

\section*{Acknowledgments}

We thank the anonymous referee for many insightful comments that greatly improved the quality of this paper.
We are very grateful to Tiffany Fields for her help with the images presented in this work, and thanks are due to Anneya Golob and Liz Arcila-Osejo for useful discussions and a careful reading of the manuscript.

This work was supported financially by the Natural Sciences and Engineering Research Council (NSERC) of Canada, including an NSERC Graduate Scholarship to RS and an NSERC Discovery Grant to MS. 

Computational facilities are provided by ACEnet, the regional high performance computing consortium for universities in Atlantic Canada. ACEnet is funded by the Canada Foundation for Innovation (CFI), the Atlantic Canada Opportunities Agency (ACOA), and the provinces of Newfoundland and Labrador, Nova Scotia, and New Brunswick. 

This research made use of Astropy, a community-developed core Python package for Astronomy \citep{Robitaille2013}, as well as many of the routines available in the scipy and numpy software packages.

Some of the data presented in this paper were obtained from the Mikulski Archive for Space Telescopes (MAST). STScI is operated by the Association of Universities for Research in Astronomy, Inc., under NASA contract NAS5-26555. Support for MAST for non-HST data is provided by the NASA Office of Space Science via grant NNX13AC07G and by other grants and contracts.

Funding for SDSS-III has been provided by the Alfred P. Sloan Foundation, the Participating Institutions, the National Science Foundation, and the U.S. Department of Energy Office of Science. The SDSS-III web site is http://www.sdss3.org/. SDSS-III is managed by the Astrophysical Research Consortium for the Participating Institutions of the SDSS-III Collaboration including the University of Arizona, the Brazilian Participation Group, Brookhaven National Laboratory, Carnegie Mellon University, University of Florida, the French Participation Group, the German Participation Group, Harvard University, the Instituto de Astrofisica de Canarias, the Michigan State/Notre Dame/JINA Participation Group, Johns Hopkins University, Lawrence Berkeley National Laboratory, Max Planck Institute for Astrophysics, Max Planck Institute for Extraterrestrial Physics, New Mexico State University, New York University, Ohio State University, Pennsylvania State University, University of Portsmouth, Princeton University, the Spanish Participation Group, University of Tokyo, University of Utah, Vanderbilt University, University of Virginia, University of Washington, and Yale University. 


\bibliographystyle{mn2e_2014}
\bibliography{pxp1}{}

\bsp

\label{lastpage}

\end{document}